# Faster than Nyquist Transmission by Non-Orthogonal Time Division Multiplexing of Nyquist Sinc Sequences

Ali Dorostkar and Thomas Schneider

**Abstract**: One possibility to break down the capacity limit in optical transmission systems are higher spectral efficiencies, enabled by Faster-than-Nyquist signaling. Here we present the utilization of non-orthogonal time division multiplexing of sinc pulse sequences for this purpose. The mathematical expression, with a representation of the Nyquist sinc sequence by a cosine Fourier series and simulation results show that non-orthogonal time-division multiplexing increases the transmittable symbol rate by up to 25%.

*Index Terms*— **Faster-than-Nyquist, sinc sequence signal, non-orthogonal time-division multiplexing**

## I. INTRODUCTION

Nowadays an increasing demand of data rate and bandwidth are taken into account for new applications such as video streaming, super high vision television or cloud computing for big data [1-2]. The Cisco Visual Networking Index Forecast declares a growing of the annual rate of internet traffic of 26% from 2016 to 2021 [3]. To keep pace with the growing data rates, optical links with higher spectral efficiencies are desired. One possibility are higher-order modulation formats and polarization multiplexing [4]. However, usually this comes at the cost of a higher complexity and energy consumption.

Faster-than-Nyquist (FTN) signaling [5-13] is an approach to provide even higher spectral efficiencies than possible with rectangular channels. Generally, FTN can be divided into two methods. The first brings the adjacent signals closer together in time domain, which results in a collapse of the orthogonality. Mazo has shown a spectral efficiency enhancement of 25% in time domain [5]. For the other method, the signal is compressed in frequency domain by squeezing the baseband bandwidth of the single carrier signal by duobinary pulse shaping filters [6-10]. In addition, the subcarrier space within a multicarrier signal can be decreased in order to further reduce the required bandwidth [11]. Besides, non-orthogonal frequency-division multiplexing (NOFDM) has been carried out by a reduction of the subcarrier spacing in comparison with an OFDM signal. An increase of the data rate by 25% has been attained by this method as well [12]. A further enhancement of the spectral efficiency can be achieved by a combination of the FTN methods in the time and frequency domain. First, the data for each carrier is accelerated in the time domain and then the carrier spacing in the frequency domain was compressed [13].

A. Dorostkar and T. Schneider are with the Institut für Hochfrequenztechnik, Technische Universität Braunschweig, 38106 Braunschweig, Germany (e-mail: ali.dorostkar@ihf.tu-bs.de, thomas.schneider@ihf.tu-bs.de)

However, the previous research for FTN has been focused either on ideal sinc pulses, or on Nyquist pulses with a roll-off factor. Ideal sinc pulses are unlimited and thus just a mathematical construct and, due to the necessary guard band, Nyquist pulses with a roll-off factor lead to a waste of bandwidth.

Thus, here we introduce a new method of FTN based on sinc pulse sequences. As will be shown, an increasing of the data rate up to 25% can be reached by non-orthogonal time division multiplexing (NOTDM) of sinc pulse sequences.

In section II we will summarize the data transmission by sinc pulse sequences, section III introduces the theory of FTN transmission with sinc pulse sequences and presents simulation results. Section IV is dedicated to the discussion of the spectral efficiency and bit error rate of the presented methods.

## II. SINC PULSE SEQUENCE TRANSMISSION

Since they show zero inter-symbol-interference (ISI) by the use of orthogonality and can even have a rectangular spectrum, Nyquist pulses take a key role for data transmission [14-16]. Thus, if used for data transmission, wavelength or frequency channels can be combined together without any guard band in between. In principle this feature is used for orthogonal frequency division multiplexing (OFDM) as well as for Nyquist time domain multiplexing. The fundamental difference between both can be seen in Fig.1. Ideal OFDM would use sinc functions in the frequency domain, leading to rectangular pulses in time. For ideal Nyquist-TDM the sinc shaped pulses are in the time domain, leading to rectangular spectra.

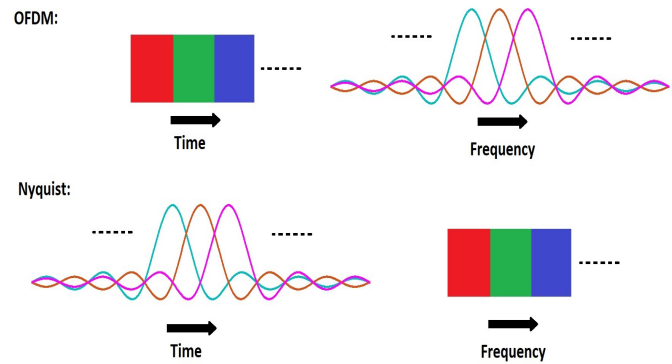

Fig.1. Comparison between OFDM (top) and Nyquist TDM (bottom) in the frequency and time domain.

With the right time-shift, sinus cardinals or sinc pulses are orthogonal to each other. Thus, each single sinc pulse can independently transmit data and this data can be retrieved from the pulse by a multiplication with a sinc pulse with the right time shift in the receiver.

Due to the orthogonality of sinc functions it can be shown that:

$$\int_{-\infty}^{+\infty} D(\frac{i}{B})\operatorname{sinc}\left(B\left(t-\frac{i}{B}\right)\right)\operatorname{sinc}\left(B\left(t-\frac{i}{B}\right)\right) = D(\frac{i}{B}) \quad (1)$$

$$\int_{-\infty}^{+\infty} D(\frac{j}{B})\operatorname{sinc}\left(B\left(t-\frac{j}{B}\right)\right)\operatorname{sinc}\left(B\left(t-\frac{i}{B}\right)\right) = 0 \quad (2)$$

with $j \neq i, j \& i \in I$ and $I$ as the set of integer numbers.

Here $D(\frac{i}{B})$ is the data at time instant $i$ and $D(\frac{j}{B})$ is the data at another time, $B$ is the bandwidth of the sinc pulse.
Thus, by a multiplication of the modulated sinc pulse with a sinc pulse at distinct time and a following integration, the data signal can be extracted. As can be seen from Eq. 2, due to orthogonality, the other modulated sinc pulses have no influence on the extracted data at the time $i$.

However, as already mentioned, ideal sinc pulses are a mathematical construct, whereas Nyquist pulses with a roll-off factor cannot be multiplexed without a guard band. This problem might be solved, however, by sinc pulse sequences.

As shown in [14], a sinc pulse sequence is the unlimited superposition of time-shifted ideal sinc pulses.

$$\sum_{n=-\infty}^{+\infty} \operatorname{sinc}\left(N\Delta f\left(t-\frac{n}{\Delta f}\right)\right) = \frac{\sin(N\pi\Delta ft)}{N\sin(\pi\Delta ft)} \quad (3)$$

Where $N\Delta f$ is the total bandwidth of the pulses with a repetition rate of $\Delta f$. The difference between a single sinc pulse and a sinc pulse sequence can be seen in Fig.2.

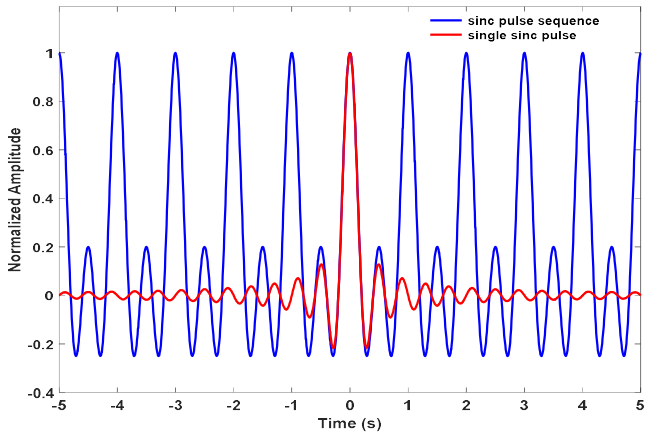

Fig. 2. comparison of the normalized amplitude for an ideal single sinc pulse (red) and a sinc pulse sequence with $N = 5$ (blue)

In complete contrast to a single sinc pulse, sinc pulse sequences can be generated by a rectangular, phase-locked frequency comb. For high-bandwidth pulses, required in optical communications, this rectangular frequency comb can be generated by one or two coupled intensity modulators, driven with one or several RF frequencies [14].
For a data transmission based on sinc sequences the data has to be modulated in amplitude or phase on the sequence over the time of the repetition rate $\Delta f$ and all these time domain channels are multiplexed together, as shown in Fig.3.

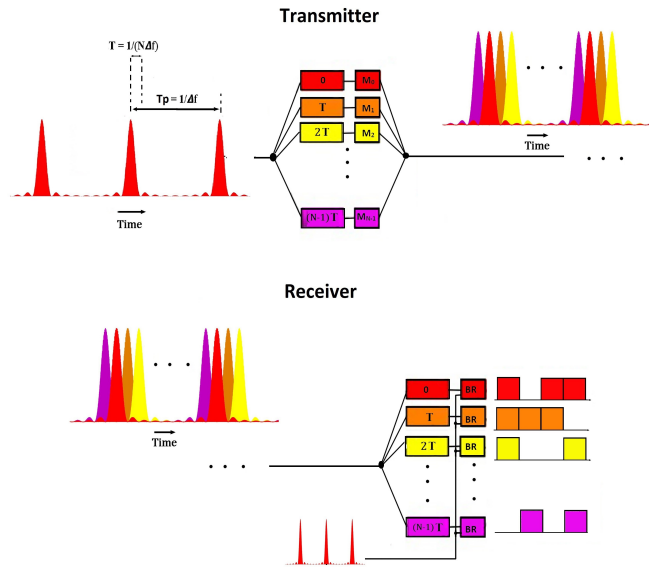

Fig.3. Basic principle of the transmitter and receiver for a data transmission with Nyquist sinc sequences.

Thus, the sinc sequences generated by intensity modulators are fed into $N$ different branches. Here $N$ is again the number of lines in the frequency comb, generated by the intensity modulators. To ensure orthogonality, the pulses of the next sequence must fall into the zero crossings of the previous one. Therefore, in each branch the sequence is delayed by a time shift equal to $\frac{n}{N\Delta f}$ and $n = 0, 1, 2, \ldots (N-1)$ as an integer.

The pulses are then modulated with the data, as shown in Fig.4. The modulation can be either an amplitude or phase modulation or each higher order modulation format.

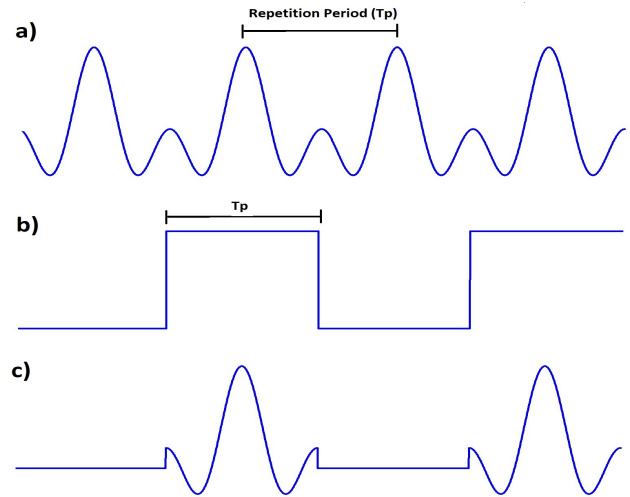

Fig.4. Nyquist sinc sequence with $N = 5$ (a), binary sequence (b) and the result after an amplitude modulation (c).

Before transmission all these sequences are multiplexed together in order to have a data rate of $N$ times the data rate of the single TDM channel. Since with low bandwidth electronics and low bandwidth modulators the data rate can be enhanced $N$ times, this method might be very interesting for integrated Silicon-on-Insulator (SOI) transceivers with their limited bandwidth.

The other basic advantage of using sinc pulse sequences is that these multiple TDM channels all together have a rectangular bandwidth. These N-TDM channels together will be called TDM-WDM channel from hereon. These TDM-WDM channels can then be multiplexed in the wavelength or frequency domain to form WDM channels without a guardband. This is shown in Fig. 5. If for the sake of simplicity, a rectangular modulation signal is assumed, this rectangular bit has the duration $T_p = 1/\Delta f$. Thus, each line of the frequency comb is multiplied with a sinc spectrum with a bandwidth from the maximum to the first zero crossing of $\Delta f$. As a result, the zero crossings of the modulated spectrum at the edges fall exactly together with the frequency lines of the next TDM-WDM channel, ensuring orthogonality between the WDM channels. However, this requires that the different center wavelength of the WDM channels have a fixed frequency relationship to each other, possible by a frequency comb for instance. That Nyquist sinc sequences can indeed be used for high bitrate communications with TDM-WDM channels without guard band, has been shown in [17] for instance.

The receiver is shown on the bottom of Fig. 3, the different TDM channels multiplexed in time domain are again divided into N-branches, each of which with the right time shift. However, for different WDM channels the process is also the same. In a balanced receiver these delayed channels are multiplied with a sinc pulse sequence with the center frequency of the required WDM channel. All other WDM channels of the incoming multiplexed signal result in a frequency beating inside the photodiode. If the bandwidth of the photodiode is low enough, this beating cannot be found in the electrical output signal. Additionally, the limited bandwidth of the photodiodes in the balanced receiver accomplishes the integration. Thus, due to the orthogonality of the sinc pulse sequences, the single TDM channel is extracted from the WDM-TDM channels.

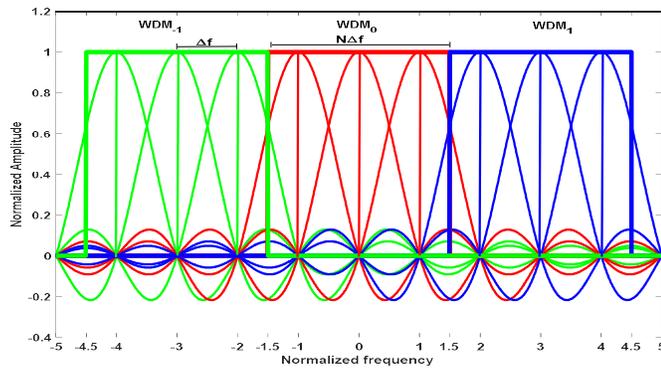

Fig. 5. WDM-TDM multiplexing of modulated sinc sequences. The modulated three TDM channels (red) build one WDM channel which is multiplexed with two adjacent WDM channels (green and blue) in the frequency domain, The frequency is normalized to the center of the $n^{th}$ channel $f_{nc}$.

As has been shown in [14 & 17], sinc pulse sequences enable the data transmission with the maximum possible symbol, or the so called Nyquist rate, which is twice the baseband bandwidth or equal to the signal bandwidth in the optical domain [12]. However, for a further increasing of the data rate, a faster than Nyquist (FTN) transmission is required.

## III. FASTER THAN NYQUIST TRANSMISSION WITH SINC PULSE SEQUENCES

FTN signals can be described as [5]:

$$y(t) = \sum_{n=-\infty}^{+\infty} b_n h(t - n\tau T), \quad (4)$$

where $y(t)$ is the transmitted signal, $b_n$ is the symbol sequence and $h(t)$ is an orthogonal pulse. Pulses shifted by $nT$ satisfy the orthogonality and have a zero ISI. For FTN the pulses must be shifted by $n\tau T$. Thus, $0 < \tau \leq 1$ can be described as an acceleration factor. In this case, the orthogonality is no longer fulfilled, resulting in an ISI.

Here instead of single, orthogonal pulses $h(t)$, a sinc pulse sequence is applied. Since the Nyquist sequence is an even periodic function, it can be written as a cosine Fourier series for an odd number of $N$ as:

$$\frac{\sin(N\pi\Delta ft)}{N\sin(\pi\Delta ft)} = \frac{1}{N} + \frac{2}{N}\sum_{n=1}^{\frac{N-1}{2}} \cos(2\pi n\Delta ft) \quad (5)$$

and for an even number of $N$ as:

$$\frac{\sin(N\pi\Delta ft)}{N\sin(\pi\Delta ft)} = \frac{2}{N}\sum_{n=1}^{\frac{N}{2}} \cos(2\pi(n-0.5)\Delta ft) \quad (6)$$

Thus, the sinc pulse sequence is a summation of cosine functions. In the frequency domain this corresponds to $N$ delta functions equally spaced by $\Delta f$, i.e. a rectangular frequency comb.

Figure 6 depicts Nyquist sinc sequences for an even and odd number of $N$ with a repetition rate of 10 GHz in the time (a) and frequency domain (b).

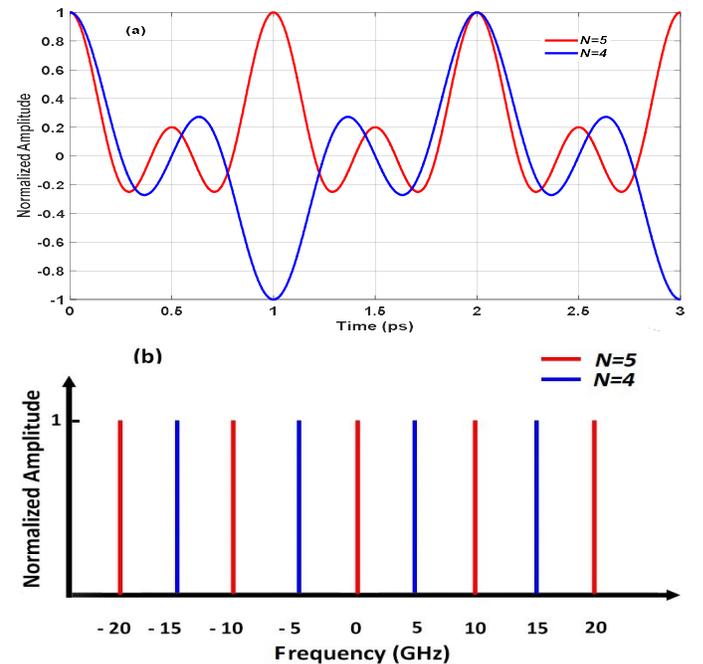

Fig.6. Nyquist sinc sequence for $N = 4$ (blue) and $N = 5$ (red) with a repetition rate of 10 GHz in time (a) and frequency domain (b). The frequency axis in (b) is normalized to the optical carrier frequency.

As proved by Mazo [5], the acceleration of single sinc pulses with a factor of $\tau = 0.8$ will not deteriorate the bit error rate (BER). Therefore, a Nyquist sinc sequence defined by a superposition of infinite sinc pulses should have the possibility for an acceleration factor of 0.8 as well. However, since it is a sequence, for higher spectral efficiencies the free space has to be filled with at least one more channel. The acceleration of the pulses with $\tau = 0.8$ generates an empty space of $0.2T$. Since 4 x 0.2 = 0.8, for a sinc sequence with $N$ = 4 this means that one new channel can be added, leading to 5 channels in the same time period. Consequently, for $N$ = 8 there is the possibility to add 2 new channels and so on. In general, for $\tau = 0.8$ the number of possible additional channels corresponds to $\left[\frac{N}{4}\right]$. Accordingly, Eq. 4 has to be changed to the following relation for Nyquist sinc sequences of $N \geq 4$:

$$y(t) = \sum_{i=0}^{N-1+\left[\frac{N}{4}\right]} \sum_{n=-\infty}^{+\infty} b_{n,i} h(t - i\tau T) \quad (7)$$

Where $i$ corresponds to the branch number of the TDM channel. For instance, $i = 0$ is the first branch with zero time delay and so on. Thus, $b_{n,0}$ is a symbol sequence for the first branch.

Fig. 7 (a) shows the Nyquist sinc sequence for $N = 4$ with the corresponding shifted versions to the zero crossings. All functions are orthogonal in one time interval of $NT$. However, as shown in Fig. 7. (b), with an acceleration factor of 0.8, one more channel can be added, but the orthogonality is lost. This will be called non-orthogonal time-division multiplexing (NOTDM) from hereon.

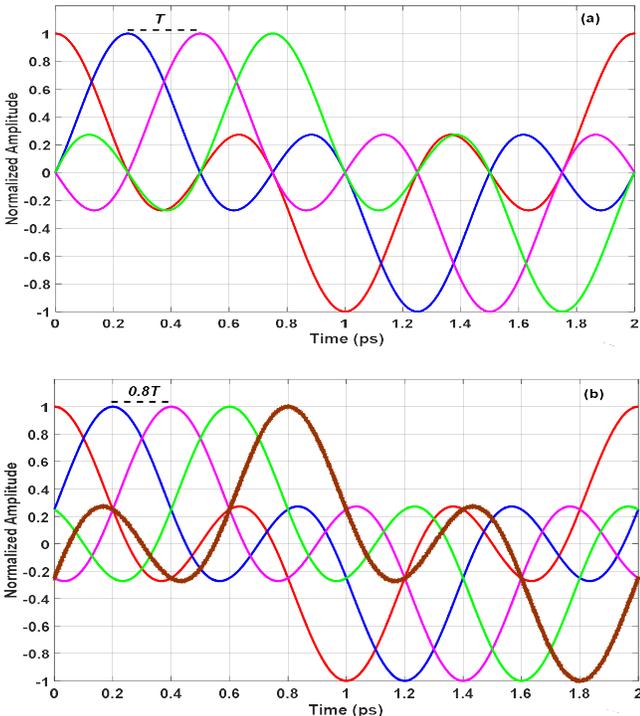

Fig. 7. OTD multiplexing of Nyquist sinc sequences in the orthogonal case (a) and in the non-orthogonal case with an acceleration factor of 0.8 (b).

A general schematic of a non-orthogonal TDM (NOTDM) link for a sinc sequence is shown in Fig. 8. The Nyquist pulse sequence is split into $N + \left[\frac{N}{4}\right]$ branches, where the signal is time delayed and the data is modulated on each channel. Afterwards they are multiplexed and transmitted.

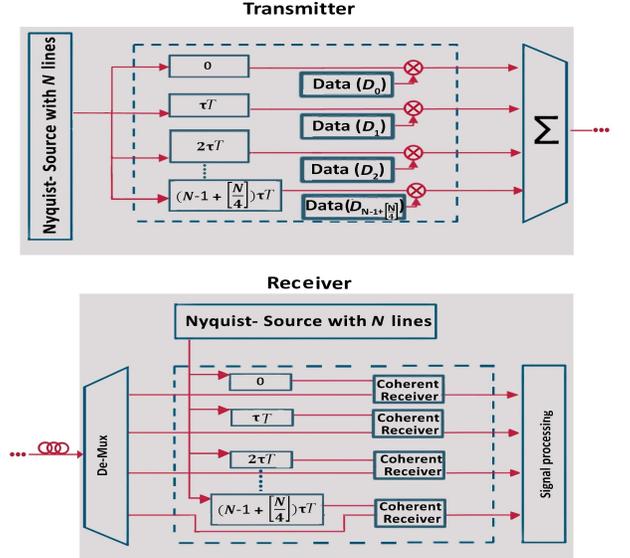

Fig. 8. Schematic of a non-orthogonal TDM communication link for sinc sequences with $N$ lines.

Within the receiver the signal is de-multiplexed and each channel is detected with a coherent receiver, where it is multiplied with an appropriate reference signal (a sinc pulse sequence with the right time-shift). The output signal of the coherent receiver for $N \geq 4$ can be expressed as:

$$R_i(t) = \int_{-\frac{1}{2\Delta f}+\frac{m}{\Delta f}}^{\frac{1}{2\Delta f}+\frac{m}{\Delta f}} \mathfrak{J}^{-1}\{H(\omega)\} * \left(\{f_i(t) \times D_i(t)\} \times f_i(t)\right) dt \quad (8)$$

Where $H(\omega)$ is the frequency response of the coherent receiver, $D_i(t)$ is a symbol sequence and $m/\Delta f$ corresponds to the $m^{th}$ symbol, $f_i(t)$ is equal to

$$\frac{1}{N} + \frac{2}{N} \sum_{n=1}^{\frac{N-1}{2}} \cos\left(2\pi n \Delta f \left(t - \frac{i\tau}{\Delta f}\right)\right)$$ and $\tau = 0.8$.

## IV. SPECTRAL EFFICIENCY AND BIT ERROR RATE

For Nyquist sequence OTDM transmission, the total symbol rate is the summation of the symbol rate of each individual channel. With $N$ channels with a symbol rate of $\Delta f$ for each channel, the total symbol rate is:

$$R_s = N\Delta f \quad (9)$$

For higher order modulation formats with $K$-bits per symbol the data rate is $KN\Delta f$. For the sake of simplicity, we have set the symbol rate equal to the data rate.

However, for the non-orthogonal condition $\left[\frac{N}{4}\right]$ additional channels for $N \geq 4$ are introduced. Accordingly, the symbol rate is:

$$R_s = (N + \left[\frac{N}{4}\right])\Delta f \quad (10)$$

Thus, for FTN in optimal condition, the symbol rate increases by $1/4$ or 25% for $N \geq 4$.

For a Nyquist signal in an additive white Gaussian noise (AWGN) channel, the Shannon capacity limit, i.e. the upper capacity limit of Nyquist signal transmission, can be calculated as [18]:

$$C \leq N\Delta f \log_2(1 + \frac{P_S}{P_N}) \quad (11)$$

Where $P_S$ and $P_N$ are the signal and noise power, respectively.

The derivation method for the capacity limit is similar to Shannon's method [19]. For the geometric method a volume of an $n$-dimensional sphere is assumed [20]. The number of distinguishable signals is the volume of the sphere of the received signals with the radius $\sqrt{(N + \left[\frac{N}{4}\right])(P_S + P_N)}$ divided by the volume of the noise perturbation sphere with a radius of $\sqrt{(N + \left[\frac{N}{4}\right])P_N}$. Therefore, the upper limit for the number of distinguishable signals is given by:

$$M \leq (\frac{P_S + P_N}{P_N})^{N + \left[\frac{N}{4}\right]} \quad (12)$$

Thus, the capacity limit can be calculated by:

$$C = \Delta f \log_2(M) \quad (13)$$

However, since for non-orthogonal TDM inter-symbol-interference occurs, the Shannon capacity limit for the proposed method can be described by:

$$C \leq (N + \left[\frac{N}{4}\right])\Delta f \log_2(1 + \frac{P_S}{P_N + P_{ISI}}) \quad (14)$$

Where the power of the ISI, $P_{ISI}$ is an additional noise term.

### A. Mapping Strategy

Noise and ISI result in an uncertainty interval for the constellation diagram at the receiver. Thus, a mapping strategy for the demodulation of the signals is required. This mapping strategy for a 2-PAM signal is depicted in Fig. 9. The correct symbol can be detected by the receiver, if it is in the yellow and blue area of Fig. 9. However, if the signal is in the uncertainty interval due to ISI, an iterative algorithm for the detection is required. Due to an increasing ISI and related uncertainty, the mapping for higher order modulation formats is more complicated. Since this requires further investigations, here we focus on 2-PAM signals.

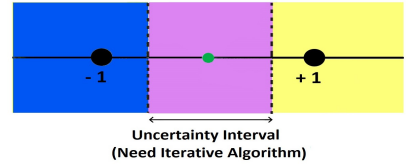

Fig. 9. Mapping strategy of 2-PAM with a symbol sequence of $\pm 1$

### B. BER Analysis

To verify the feasibility of NOTDM transmission, the BER analysis is carried out by MATLAB Simulink with the mapping strategy shown in Fig. 9. Nyquist pulses for $N = 4$ with a repetition rate of 10 GHz are simulated for a 2-PAM modulation ($\pm 1$) with a sequence consisting of $2^{18}$ symbols. The bit error rate for the binary sequence can be calculated under the best detection condition to [5]:

$$BER = Q(\frac{d_{min}^2 E_b}{N_0}) \quad (15)$$

Where $Q$ is a complementary-error function, $E_b$ is the bit energy and $N_0$ is the power spectral density of the AWGN. The minimum Euclidian distance $d_{min}$ presented in Eq. 15 is [5]:

$$d_{min}^2 = \frac{1}{2E_b} \frac{1}{T} \int_{-\frac{T}{2}}^{\frac{T}{2}} |S_i(t) - S_j(t)|^2 dt \quad (16)$$

The calculated BER versus signal-to-noise ratio ($E_b/N_o$) of sinc sequences with $N = 4$ for the orthogonal condition with an acceleration factor of $\tau = 1$, and for the non-orthogonal condition with $\tau = 0.8$, for 2-PAM modulation are depicted in Fig. 10. Due to the acceleration factor, one more channel can be added but the resulting ISI leads to more mapping data in the uncertainty interval and the BER increases. However, by iterative algorithms this drawback can be eliminated [12].

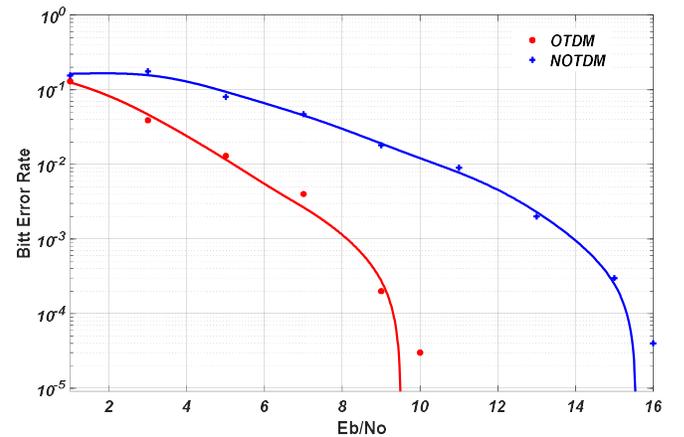

Fig. 10. BER versus $E_b/N_0$ of a sinc sequence with $N = 4$ for OTDM (red) and NOTDM (blue).

The collapse of orthogonality leads to an ill condition for an NOTDM system. However, there are some mathematical methods for recovering ill conditioned systems [21-22]. In general, the data extraction with a small BER needs a high signal to noise ratio (SNR) and low SNR values need more signal processing, e.g. by iterative detection [12]. However, in order to reach a lower BER, for NOTDM with sinc sequences new iterative algorithms have to be investigated further.

### V. Conclusion:

A new scheme for FTN transmission by the use of Nyquist sinc sequences is investigated. Contrary to sinc pulses, sinc sequences can be generated quite easily and in contrast to Nyquist pulses with a roll-off factor, no guard band is required for their multiplexing in the frequency domain. For $N \geq 4$ the symbol rate enhancement can reach 25%. However, this increases the ISI noise and for a BER of $10^{-5}$ it requires a 6 dB higher $E_b/N_o$. The BER deterioration due to the ISI can be improved by the utilization of iterative algorithms. However, this requires further investigation. The FTN transmission with NOTDM signals is a very promising method for ultra-high data rate optical communications, but can be used in other systems, like wireless links as well.

**Funding and Acknowledgments**

The authors acknowledge the financial support from the Volkswagen Stiftung (GZ 21.1-76251-87150700); and the Deutsche Forschungsgemeinschaft (DFG) (SCHN 716/13-1). Additionally, the authors like to thank Stefan Preussler and Naghmeh Akbari from the TU-BS for their support during the writing of the paper.